\documentclass[fleqn,usenatbib]{mnras}

\usepackage{newtxtext,newtxmath}
\usepackage[T1]{fontenc}

\DeclareRobustCommand{\VAN}[3]{#2}
\let\VANthebibliography\thebibliography
\def\thebibliography{\DeclareRobustCommand{\VAN}[3]{##3}\VANthebibliography}

\usepackage{graphicx}	% Including figure files
\usepackage{amsmath}	% Advanced maths commands
\title[Shocks and self-gravity in TDE streams]{On the relative importance of shocks and self-gravity in modifying tidal disruption event debris streams}

\author[Fancher et al.]{
Julia Fancher,$^{1}$\thanks{E-mail: jlfanche@syr.edu}
Eric.~R.~Coughlin,$^{1}$
and C.~J.~Nixon$^{2}$
\\
$^{1}$Department of Physics, Syracuse University, Syracuse, NY 13210, USA\\
$^{2}$School of Physics and Astronomy, Sir William Henry Bragg Building, Woodhouse Ln., University of Leeds, Leeds LS2 9JT, UK
}

\date{Accepted XXX. Received YYY; in original form ZZZ}

\pubyear{2023}

\begin{document}
\label{firstpage}
\pagerange{\pageref{firstpage}--\pageref{lastpage}}
\maketitle

\begin{abstract}
In a tidal disruption event (TDE), a star is destroyed by the gravitational field of a supermassive black hole (SMBH) to produce a stream of debris, some of which accretes onto the SMBH and creates a luminous flare. The distribution of mass along the stream has a direct impact on the accretion rate, and thus modeling the time-dependent evolution of this distribution provides insight into the relevant physical processes that drive the observable properties of TDEs. Analytic models that only account for the ballistic evolution of the debris do not capture salient and time-dependent features of the mass distribution, suggesting that fluid dynamical effects significantly modify the debris dynamics. Previous investigations have claimed that shocks are primarily responsible for these modifications, but here we show -- with high-resolution hydrodynamical simulations -- that self-gravity is the dominant physical mechanism responsible for the anomalous (i.e., not predicted by ballistic models) debris stream features and its time dependence. These high-resolution simulations also show that there is a specific length scale on which self-gravity modifies the debris mass distribution, and as such there is enhanced power in specific Fourier modes. Our results have implications for the stability of the debris stream under the influence of self-gravity, particularly at late times and the corresponding observational signatures of TDEs.
\end{abstract}

\begin{keywords}
black hole physics -- galaxies: nuclei -- hydrodynamics -- transients: tidal disruption events
\end{keywords}

%%%%%%%%%%%%%%%%%%%%%%%%%%%%%%%%%%%%%%%%%%%%%%%%%%

%%%%%%%%%%%%%%%%% BODY OF PAPER %%%%%%%%%%%%%%%%%%

\section{Introduction}

When a star passes within a critical distance of a supermassive black hole (SMBH), it is destroyed by the tidal field -- the difference in the gravitational field of the black hole across the diameter of the star -- and transformed into a stream of debris \citep[][see \citealt{gezari21} for a recent review]{hills75, rees88}. This critical distance is typically defined in terms of the tidal radius $r_{\rm t} = R_{\star}(M_{\bullet}/M_{\star})^{1/3}$, where $M_{\bullet}$ is the mass of the black hole and $R_{\star}$ and $M_{\star}$ are the stellar radius and mass, respectively, which arises from equating the stellar surface gravitational field to the tidal acceleration (and dropping order-unity numerical factors). Numerical simulations have shown that, depending on the star's internal structure and specifically its density profile, the precise distance at which the star is completely destroyed can be slightly larger than or a factor of a few smaller than $r_{\rm t}$ \citep{guillochon13, mainetti17, golightly19, lawsmith20, nixon21}. For a TDE in which the star is originally on a parabolic (or near-parabolic) orbit about the SMBH, approximately half of the stellar debris is bound to the SMBH and returns to the point of disruption, where it is expected to circularize into an accretion disc\footnote{Though the nature of this disc could be very different from a standard, thin disc \citep{shakura73, pringle81} given the high accretion rates and low binding energy of the returning debris \citep{loeb97, coughlin14, roth16, dai18, steinberg22, metzger22, sarin23}, which could help to explain the lower-than-expected temperatures seen from many TDEs; see the discussion regarding TDE optical emission in \citet{gezari21} and references therein.} (e.g., \citealt{hayasaki13, shiokawa15, bonnerot16, sadowski16, curd19, andalman22}) and viscously liberate energy at a rate comparable to the Eddington limit of the SMBH \citep{evans89, wu18}, and we are now detecting many TDEs with surveys such as ZTF \citep{bellm19}, ASAS-SN \citep{shappee14}, ATLAS \citep{tonry18}, eROSITA \citep{predehl21}, and (soon) the Rubin Observatory/LSST \citep{ivezic19} (for recent observations of TDEs, see, e.g., \citealt{komossa99, esquej07, gezari09, bloom11, cenko12, gezari12, holoien14, miller15, alexander16, cenko16, holoien16, kara16, vanvelzen16, blanchard17, gezari17, hung17, pasham17, brown18, pasham18, blagorodnova19, nicholl19, pasham19, saxton19, wevers19, hung20, hinkle21, vanvelzen21, payne21, wevers21, lin22, nicholl22, hammerstein23, pasham23, wevers23, yao23}).

An analytical model for describing the dynamics of the debris from a TDE, known as the frozen-in approximation and originally due to \citet{lacy82} (see also \citealt{bicknell83, stone13}), treats the interaction between the star and the SMBH as occurring instantaneously as the star crosses the tidal sphere of the SMBH. Specifically, at distances much greater than $r_{\rm t}$, the tidal force from the SMBH is assumed to be sufficiently weak that the star retains exact hydrostatic equilibrium, with all of the stellar material moving with the center of mass. Once the star passes within $r_{\rm t}$, the star is ``destroyed,'' implying that the stellar material moves ballistically in the gravitational field of the SMBH with each gas parcel following an independent Keplerian (or the relativistic generalization) orbit.

The frozen-in approximation predicts that the specific energy of the debris is established at the time the star passes through $r_{\rm t}$ (not the pericenter distance, which may be considerably smaller than $r_{\rm t}$, as written in, e.g., \citealt{evans89, ulmer99, lodato09, strubbe09}). Consequently, material with a negative specific energy (or a positive binding energy) -- being half of the star in this case -- is bound and returns to the SMBH at a rate proportional to $t^{-5/3}(dM/d\epsilon)$ \citep{rees88, phinney89}, where $t$ is approximately time since disruption and $dM/d\epsilon$ is the differential amount of mass at a given specific Kepelerian binding energy $\epsilon$. Since the specific binding energy is conserved in a Keplerian potential, the amount of mass per unit energy, $dM/d\epsilon$, is a function only of $\epsilon$, and one can use the energy-period relationship for a Keplerian orbit to write $dM/d\epsilon(t)$, thereby generalizing the result of \cite{rees88} that adopted a constant value for $dM/d\epsilon$. \cite{lodato09} performed this generalization, and showed that the specific energy distribution could be predicted as a function of the progenitor’s density profile (assumed unaltered until the tidal radius is reached). From their model (see also the black, dashed curves in Figures \ref{fig:1M_dmde} and \ref{fig:latertimes} below), the stellar debris is expected to have a smooth $dM/d\epsilon$ curve that is maximized and symmetric about zero, and extends from $-\Delta \epsilon$ to $+\Delta \epsilon$, where $\Delta \epsilon = GM_{\bullet}R_{\star}/r_{\rm t}^2$ is the maximum specific binding energy to the SMBH. However,  \cite{lodato09} also ran hydrodynamical simulations of TDEs, and found that their numerically obtained distributions had sharper features than their model predicted, specifically what they termed ``wings'' that had more mass at larger values of the specific energy than expected. They attributed these discrepancies and additional features in the $dM/d\epsilon$ curve to shocks in the tidal tails of the stellar debris.

However, the importance of shocks on the dynamics of the debris has recently been questioned by \cite{norman21, coughlin22, kundu22}, who found that even for deep TDEs, in which the stellar pericenter distance is well within the canonical tidal radius, shocks during the compression of the star were either weak with a Mach number near unity or absent completely\footnote{It seems likely that a shock forms near the surface following the compression and rebound of the star in deep TDEs, but this only impacts the outer stellar layers \citep{kobayashi04, guillochon09, yalinewich19}}. Contrarily, it seems possible that the other physical effect ignored in the frozen-in approximation -- the self-gravity of the debris stream -- could be responsible for the deviation of the numerically obtained $dM/d\epsilon$ curve from the analytic prediction. Various works have suggested that self-gravity can play an important role in modifying the dynamics of the debris \citep{kochanek94, guillochon14, coughlin16, steinberg19} and its hydrodynamic stability \citep{coughlin15, coughlin23}, and in the extreme case of a partial TDE in which only the outer envelope of the star is tidally stripped, the gravitational influence of the core (i.e., self-gravity) results in a departure of the asymptotic fallback rate from the expected $\propto t^{-5/3}$ scaling \citep{guillochon13} to $\propto t^{-9/4}$ \citep{coughlin19, miles20, nixon21}.

Here we directly assess and isolate the effects of self-gravity vs.~pressure (i.e., shocks) on the evolution of the debris stream from a TDE and the corresponding fallback rate. In Section 2 we present the results of numerical hydrodynamical simulations of a TDE with and without self-gravity and the resulting evolution of the specific energy distribution or lack thereof, and we conclude in Section 3.

\section{Numerical Simulations}
\label{sec:numerical}
\subsection{Setup}
\label{sec:setup}
We used the smoothed particle hydrodynamics (SPH) code \textsc{phantom} \citep{price18} to simulate the tidal disruption of a solar-type star (where $M_{\star}=M_{\odot}$ and $R_{\star}=R_{\odot}$) by a $10^6 M_{\odot}$ SMBH. The star is initially assumed to have a polytropic equation of state with adiabatic index $\gamma = 5/3$ and is relaxed by allowing the star to evolve in isolation after the polytrope with the desired density profile has been constructed. This lets the internal properties of the star reach equilibrium before it encounters the SMBH. The relaxed star is placed at a distance of $5r_{\rm t}$ from the black hole with the center of mass on a parabolic orbit where the distance from the hole at pericenter is equal to the tidal radius. The gas retains its adiabatic equation of state throughout the TDE. For additional details on the specifics of the SPH kernel and the implementation of self-gravity, see \cite{coughlin15} and \cite{price18}.

We ran two sets of simulations with identical initial parameters. In the first set of simulations, self-gravity is included throughout the TDE; in the second, self-gravity is no longer included after the star reaches pericenter. Simulations were performed with a range of particle numbers from 250 thousand to 128 million to assess convergence.

\subsection{Results}
\label{sec:results}

\begin{figure}
  \includegraphics[width=0.495\textwidth]{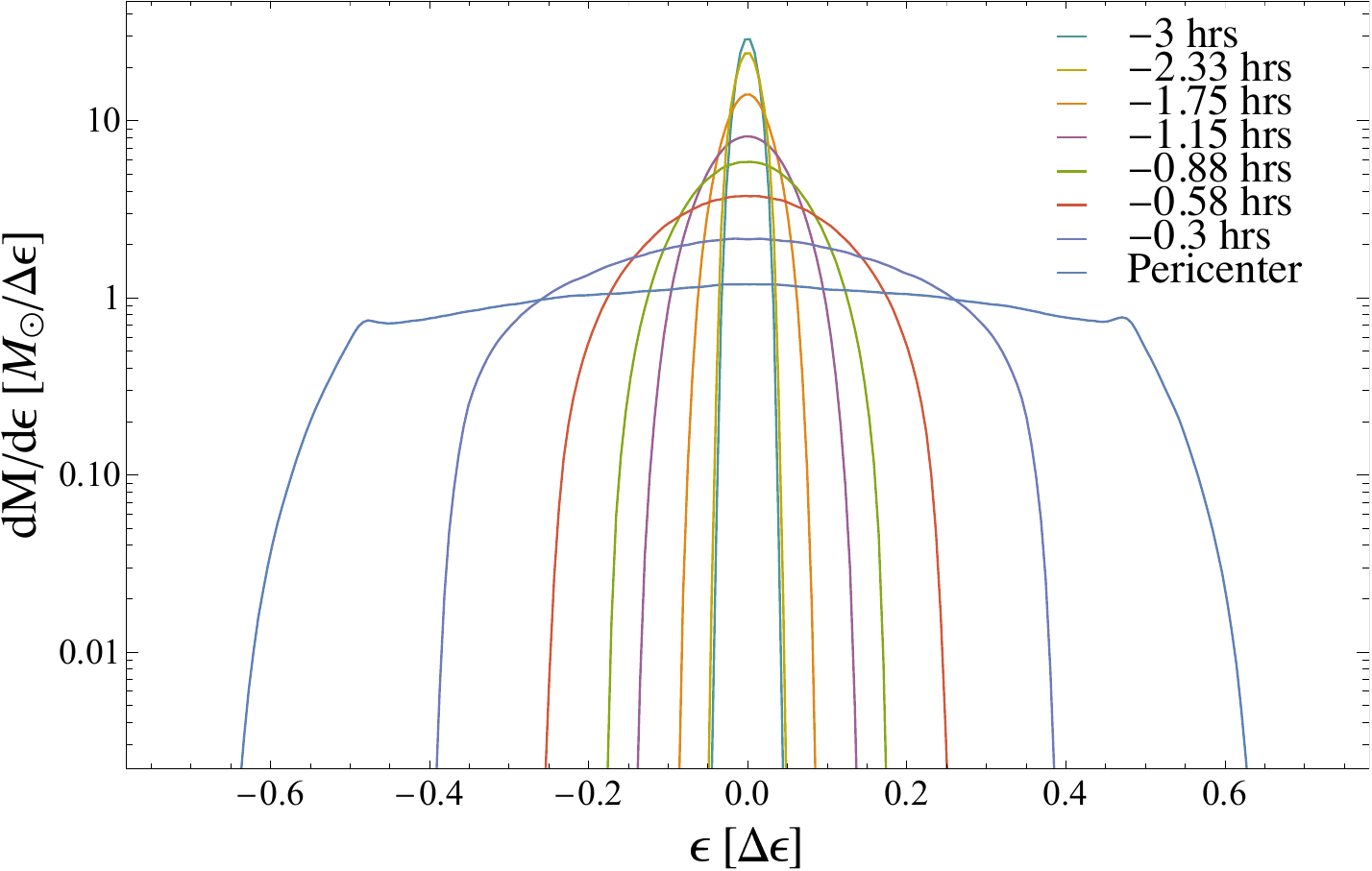}
  \caption{{The specific energy distribution of the star as it approaches pericenter, calculated from the SPH simulation with 128 million particles. The first time of $-3$ hrs (the minus sign indicates that the time is before pericenter) is when the star is at its initial distance of $5r_{\rm t}$ from the black hole. We see that the energy spread changes continuously and gradually as the star approaches the black hole, in agreement with \citet{steinberg19}.}}
    \label{fig:early}
\end{figure}

{Figure \ref{fig:early} shows the differential amount of mass per unit energy, $dM/d\epsilon$, as a function of the specific Keplerian binding energy, $\epsilon$, as the star approaches pericenter (i.e., during its initial ingress through the tidal sphere). The curves are calculated from the SPH simulation with 128 million particles, and we used 500 energy bins evenly spaced between $-1.5\Delta\epsilon$ and $1.5\Delta \epsilon$, where $\Delta\epsilon = GM_{\bullet}R_{\odot}/r_{\rm t}^2$ is the canonical energy spread from the frozen-in approximation. The legend gives the times at which the energy distributions are calculated, where the minus sign indicates that the star has not yet reached pericenter, and at $t = -3$ hrs the star is at a distance of $5 r_{\rm t}$ from the black hole. For this simulation, self-gravity is included until the star reaches pericenter. In agreement with the results found by \citet{steinberg19}, we see that the spread in the binding energies of the debris occurs over an extended period of time, reaching a value that is comparable to $\Delta \epsilon$ when the star is near pericenter. Thus, the approximation that the energy distribution is instantaneously set as the star passes through the tidal radius is rather crude.}

\begin{figure*}
    \includegraphics[width=0.495\textwidth]{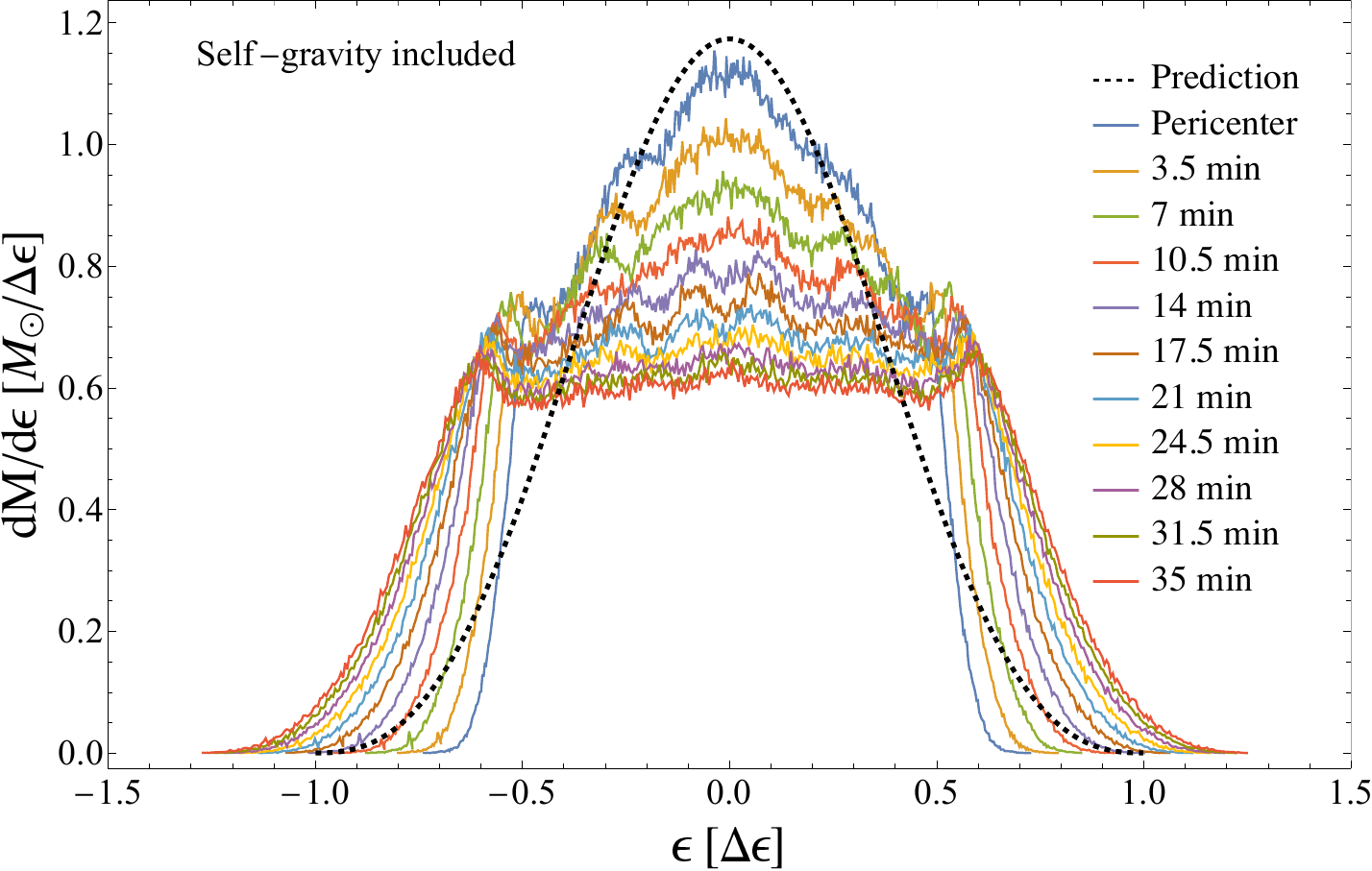}
     \includegraphics[width=0.495\textwidth]{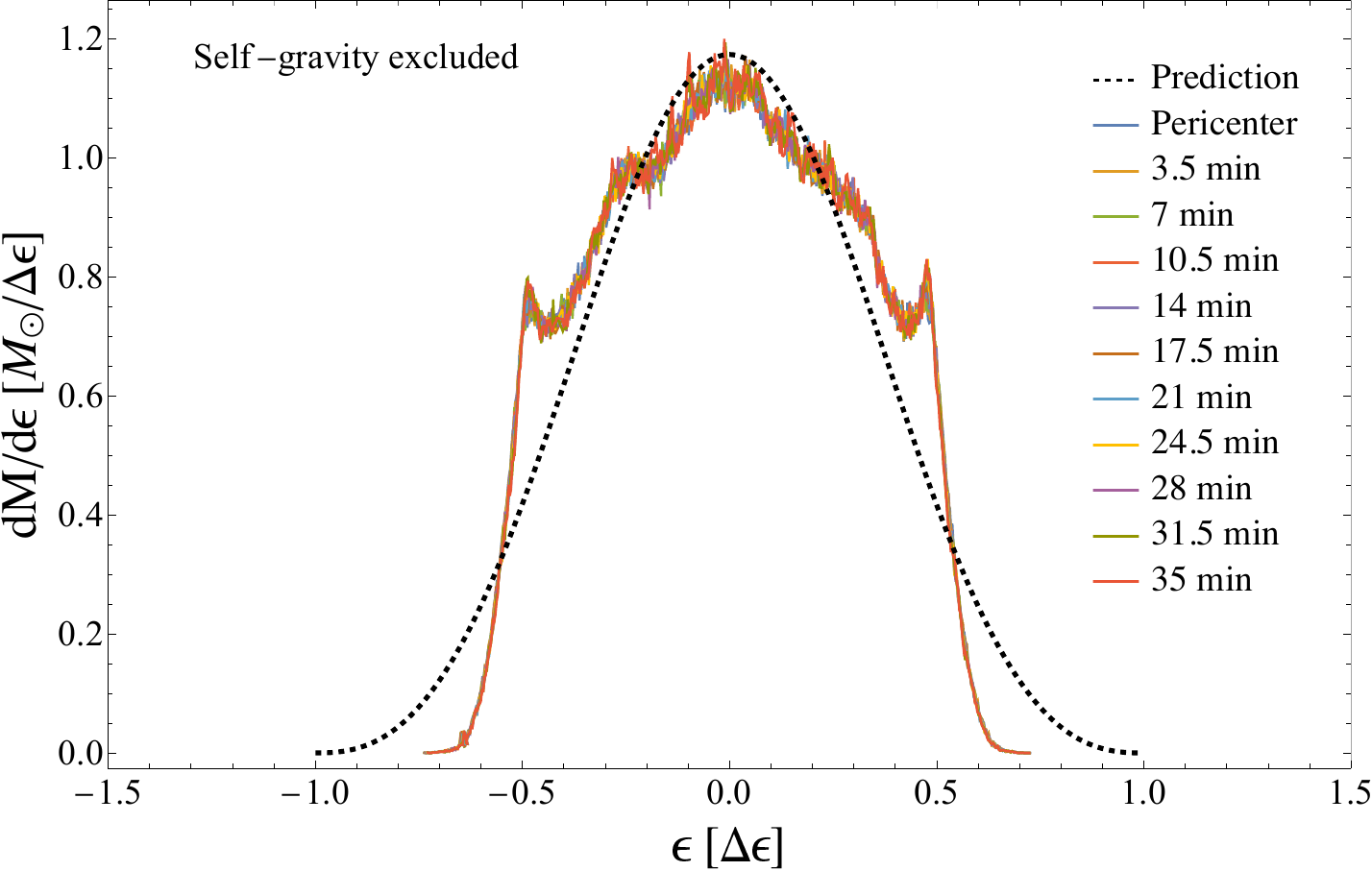}
    \caption{The evolution of the specific energy distribution with (left) and without self-gravity (right; self-gravity was turned off at pericenter) at a resolution of 1 million particles. The dashed line shows the predicted $dM/d\epsilon$ curve calculated from the frozen-in approximation. The approximation is a good fit for the data near pericenter in both cases, but becomes less accurate as the stream evolves with self-gravity.}
    \label{fig:1M_dmde}
\end{figure*}

The left panel of Figure \ref{fig:1M_dmde} shows the $dM/d\epsilon$ curves calculated every 3.5 minutes after pericenter from the simulation with 1 million particles and self-gravity included. The predicted $dM/d\epsilon$ curve indicated by the dashed black line is calculated from the frozen-in approximation as in \cite{lodato09}. The prediction approximately matches the simulated curve at pericenter (shown in dark blue), but the curves from the simulation immediately evolve to become wider and flatter, making the prediction inaccurate within minutes from disruption. The $dM/d\epsilon$ curves from the 1 million particle simulation without self-gravity are shown in the right panel of Figure \ref{fig:1M_dmde}, where the dashed black line is again the prediction. Without self-gravity, the $dM/d\epsilon$ curve is essentially frozen-in at pericenter, implying that pressure is dynamically insignificant. 

\begin{figure}
  \includegraphics[width=0.495\textwidth]{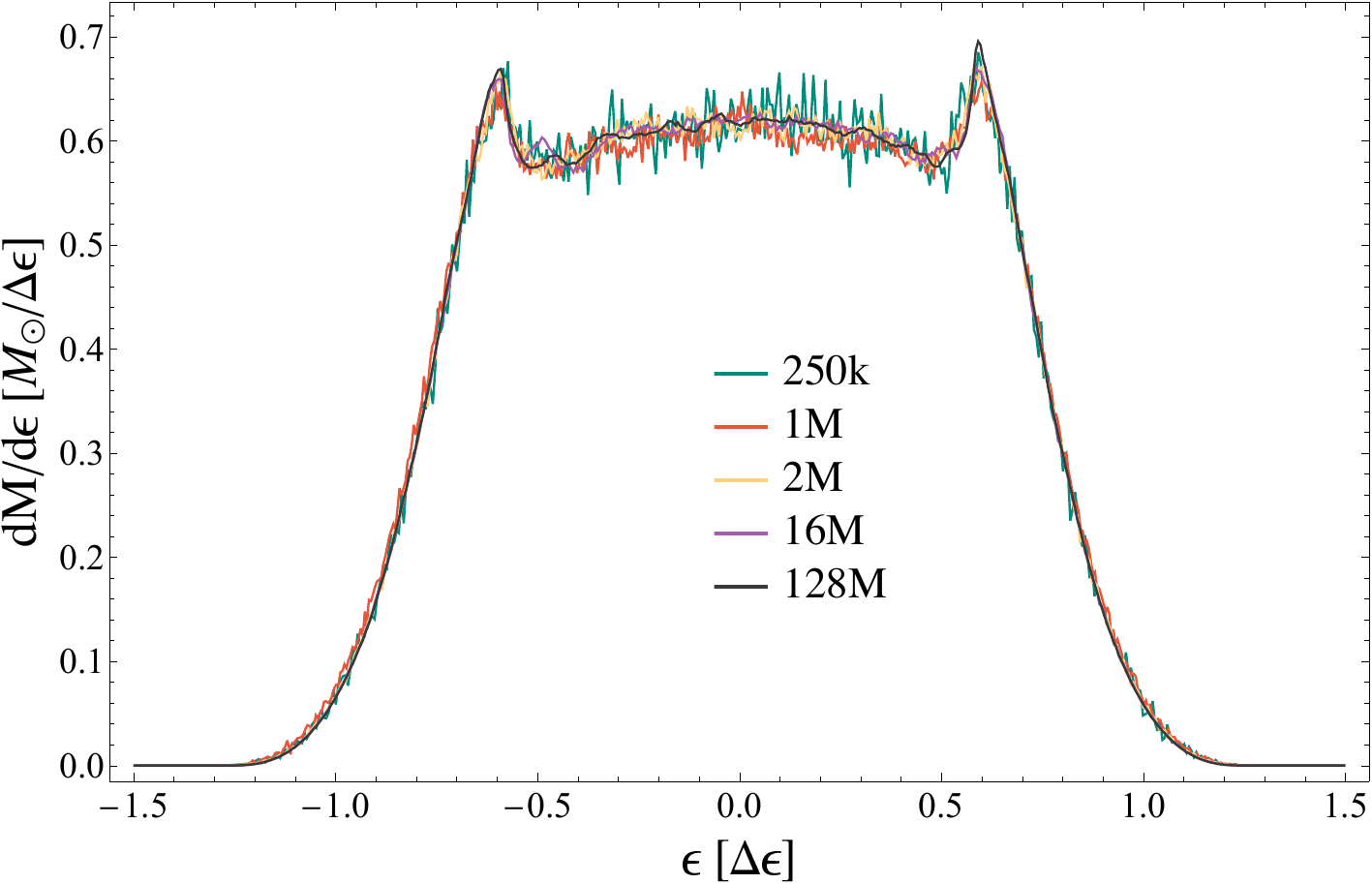}
  \caption{The $dM/d\epsilon$ curve 35 minutes after pericenter at different resolutions. Note that there are still short-lengthscale variations along the debris stream even at high resolutions, which suggests there are physical density oscillations in the disrupted star.}
    \label{fig:overlay}
\end{figure}

Both sets of $dM/d\epsilon$ curves in Figure \ref{fig:1M_dmde} have large fluctuations between $-\Delta\epsilon/2$ and $\Delta\epsilon/2$. To assess whether this feature is related to Poisson noise or physical, we plotted the $dM/d\epsilon$ curve calculated 35 minutes after pericenter from simulations with self-gravity included at different resolutions, shown in Figure \ref{fig:overlay}. It is clear that at lower resolution the amplitude and the number of fluctuations increases, but above 16 million particles the curves are relatively smooth and resemble each other. This implies that while there is some numerical noise related to particle binning, there are underlying physical oscillations in the density of the disrupted stream.

\begin{figure}
  \includegraphics[width=0.495\textwidth]{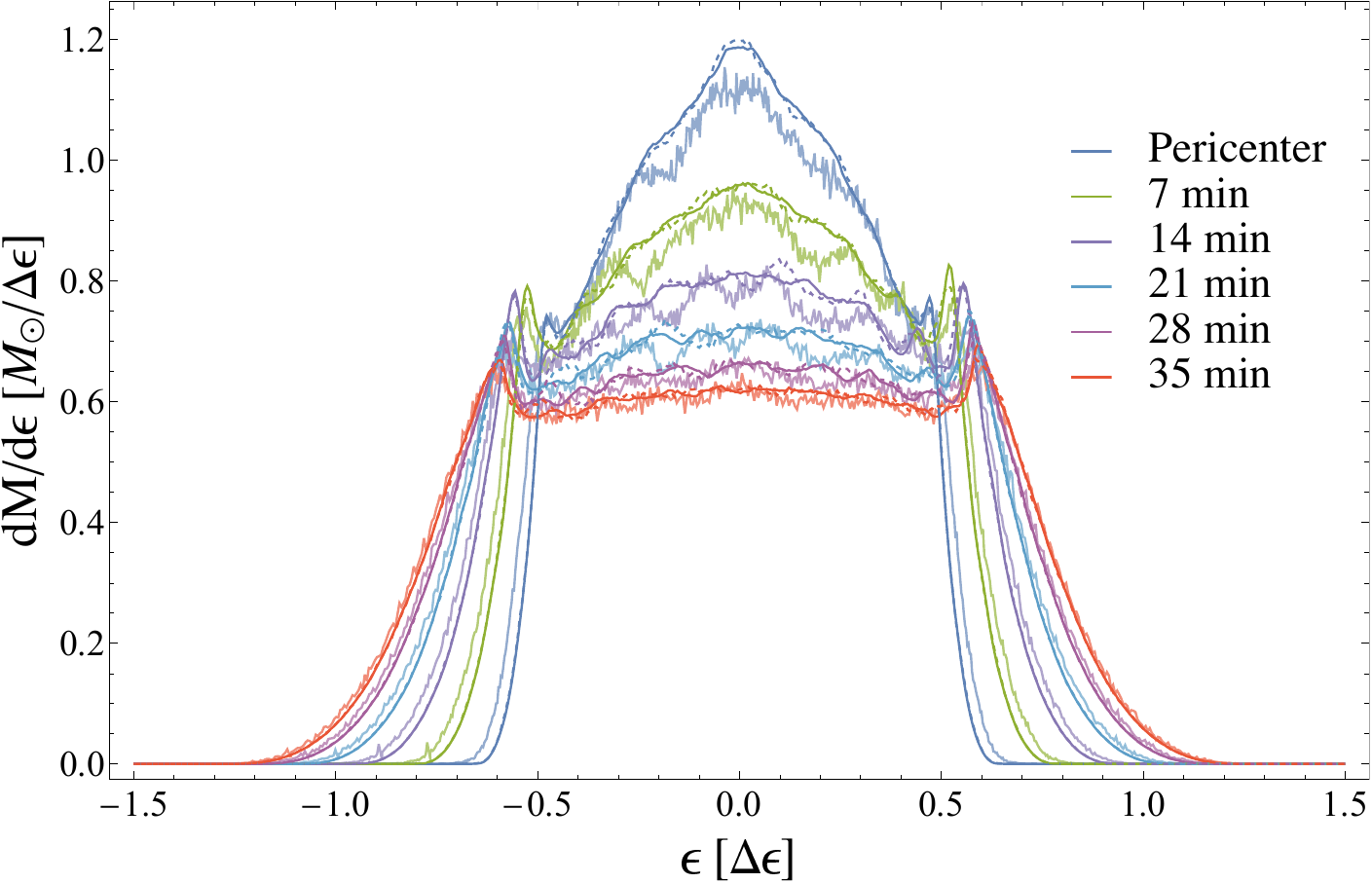}
    \caption{The $dM/d\epsilon$ curves with self-gravity at different resolutions. The lighter solid lines correspond to 1 million particles (Figure \ref{fig:1M_dmde}), the dashed lines to 16 million particles, and the dark solid lines to 128 million particles.}
    \label{fig:100M_dmde}
\end{figure}

To further assess the convergence of the simulations, in Figure \ref{fig:100M_dmde} we over-plotted the $dM/d\epsilon$ curves from the 16 million (dashed lines) and 128 million (dark solid lines) particle simulations with the $dM/d\epsilon$ curves from the 1 million particle simulation (light solid lines) as seen in Figure \ref{fig:1M_dmde}. The colors of the curves match the colors in Figure \ref{fig:1M_dmde} for the same times. The approximately sinusoidal fluctuations seen between $14-21$ minutes in Figure \ref{fig:1M_dmde} are still apparent at the 128 million particle resolution, although the feature is now strongest between $21-28$ minutes. At all resolutions, this feature is much weaker but still present at 35 minutes.

\begin{figure}
  \includegraphics[width=0.495\textwidth]{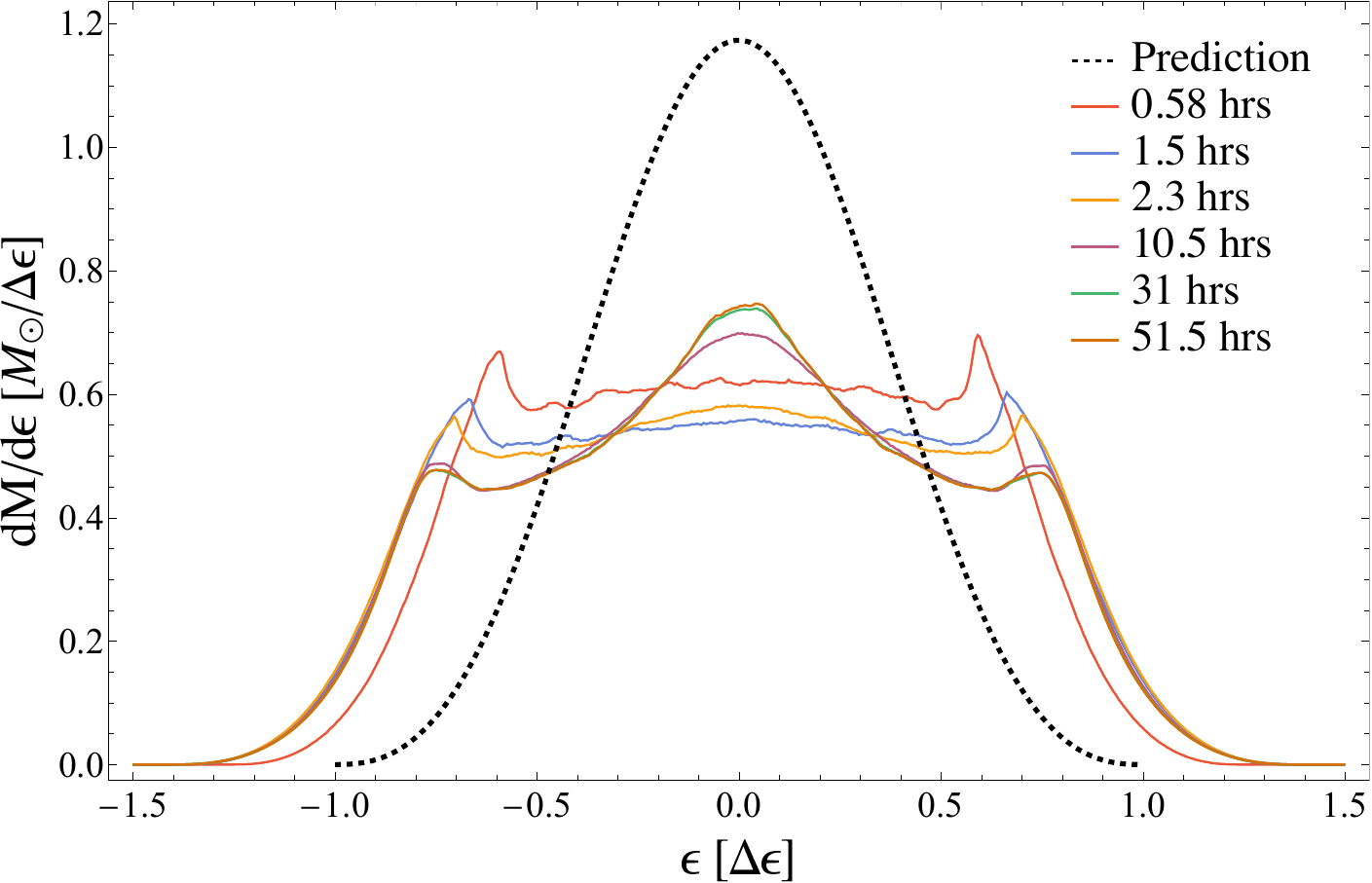}
    \caption{The $dM/d\epsilon$ curves with 128 million particles at later times past pericenter, during which the central peak reforms. After the curve at 51.5 hours, the $dM/d\epsilon$ distribution remains essentially constant for the next 66 days, which marks the end of the simulation.}
    \label{fig:latertimes}
\end{figure}

Figure \ref{fig:latertimes} shows the the $dM/d\epsilon$ curves from later times with self-gravity. The $dM/d\epsilon$ curve continues to flatten until $\sim 1.5$ hours after pericenter, after which time it ``bounces'' and the central peak begins to reform, suggesting material is coalescing toward the marginally bound radius. By this time, the sinusoidal variations are no longer visibly present, and the curves are much smoother than those around pericenter. Note that the difference in time between successive curves is much greater than that adopted in Figure \ref{fig:1M_dmde}, and it is only on these much longer timescales that the $dM/d\epsilon$ curves show temporal variability. This slowing of the debris stream evolution is consistent with the findings of  \cite{coughlin23}, who found that perturbations on top of the background state of the stream -- which is over-stable -- vary logarithmically with time. The $dM/d\epsilon$ curve remains essentially constant from the curve at 51.5 hours (seen in Figure \ref{fig:latertimes}) until the end of the simulation, which runs to approximately 68 days. The $dM/d\epsilon$ curve for later times is wider than the curve at pericenter, and the central peak at later times never reaches the height of the peak at pericenter, meaning the prediction from the frozen-in approximation is inaccurate at later times for both the height of the peak and range of $\Delta\epsilon$ (note that \citealt{lodato09} arbitrarily rescaled the height of their frozen-in model to match their numerically obtained $dM/d\epsilon$ curve at late times, meaning, among other things, that the mass of their frozen-in model did not equal that of their numerical simulation).

\begin{figure}
  \includegraphics[width=0.44\textwidth]{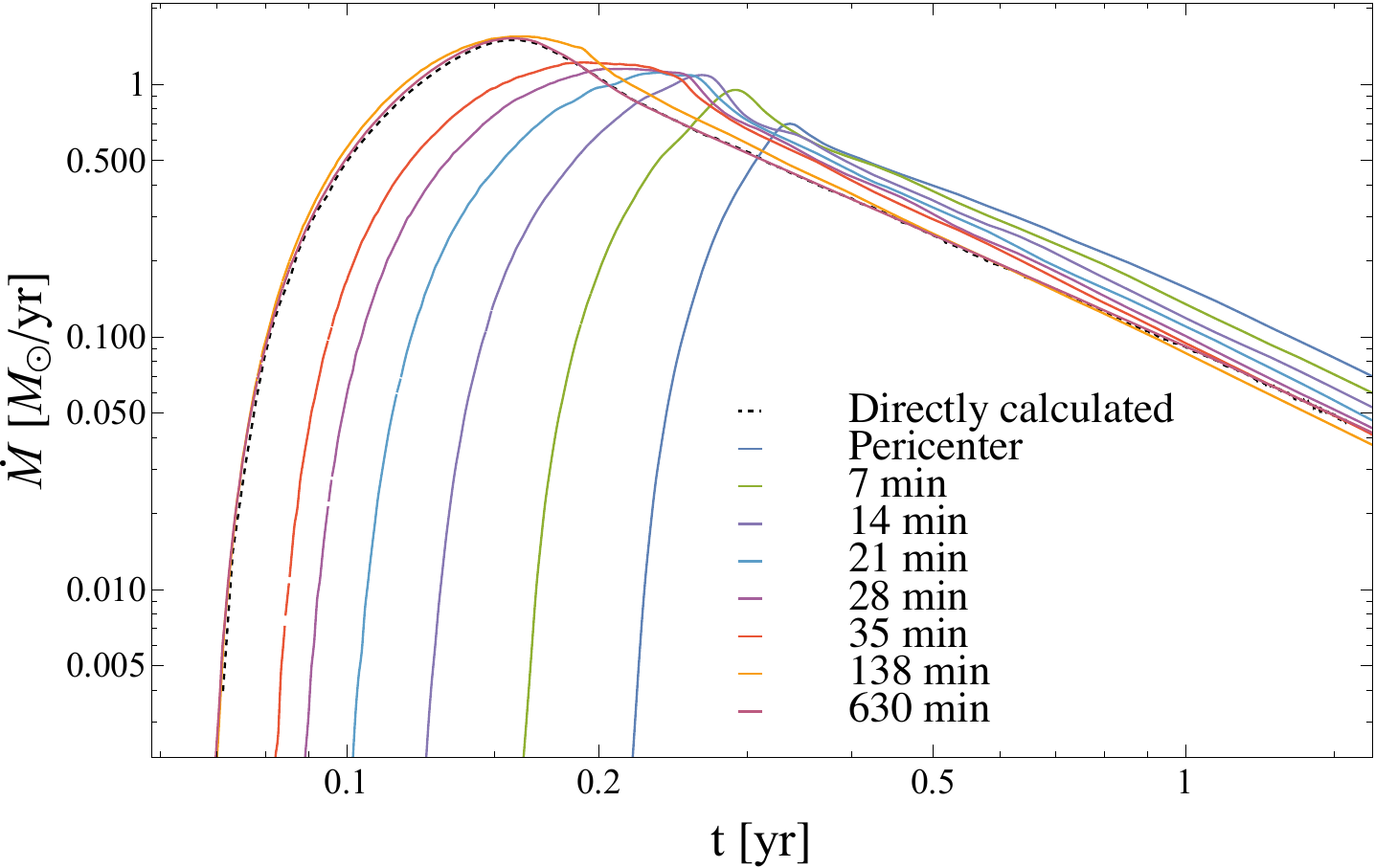}
  \includegraphics[width=0.5\textwidth]{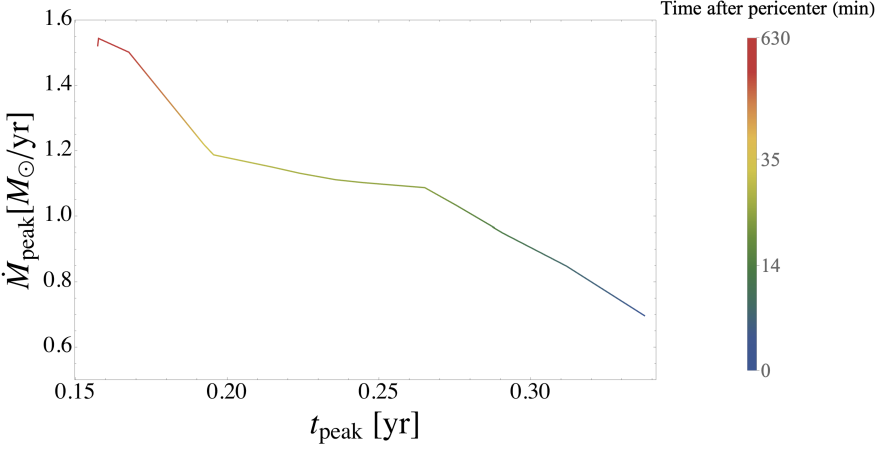}
    \caption{Top: The fallback rates calculated from the $dM/d\epsilon$ curves at the times specified in the legend. The dashed line indicates the fallback rate directly calculated from the 16 million particle run. Bottom: The evolution of the predicted peak fallback rate (calculated from $dM/d\epsilon$) as a function of the time after disruption when the peak is expected to occur. The shading indicates the time at which the fallback rate was calculated, with early times on the right in blue.}
    \label{fig:fb}
\end{figure}

The $dM/d\epsilon$ curves can be used to calculate the fallback rate, $\dot{M}$, at a given time after disruption according to \citep{lodato09, guillochon13}
\begin{equation}
    \dot{M}=\frac{(2\pi G M_{\bullet})^{2/3}}{3} \frac{dM}{d\epsilon} t^{-5/3},
\end{equation}
where $M_{\bullet}$ is the mass of the black hole and
\begin{equation}
    \epsilon(t) = -\frac{1}{2} \left( \frac{2\pi G M_{\bullet}}{t} \right) ^{2/3}.
\end{equation}
 The evolution of the $dM/d\epsilon$ curve when self-gravity is included therefore results in a changing fallback rate that is not predicted when using the frozen-in approximation (see also the discussion in \citealt{guillochon13}); {\citet{wu18} showed that the canonical scaling of the fallback rate as $\propto M_{\bullet}^{1/2}$, as discussed in \citet{lacy82} and \citet{rees88}, is accurately recovered for black hole masses between $10^5 M_{\odot} \le M_{\bullet} \le 10^{7} M_{\odot}$; see specifically their Figure 1}.\footnote{{For either extremely large or extremely small black hole masses, the behavior shown in Figure \ref{fig:fb} must change qualitatively, as in the large-mass limit relativistic effects will introduce an additional mass dependence, while in the small-mass limit, the fallback time will not be much longer than the dynamical time of the original star. Since self-gravity acts on the latter timescale (as shown in Figure \ref{fig:100M_dmde}), if the fallback time is shorter than the dynamical time, self-gravity will not have enough time to modify the energy distribution. Figure \ref{fig:latertimes} illustrates that the energy distribution does not evolve significantly after $\sim$ 10 hours; setting 10 hours equal to $T_{\rm  fb}$, where $T_{\rm fb} = 2\pi \left(R_{\star}/2\right)^{3/2}/\sqrt{GM_{\star}}
\times (M/M_{\star})^{1/2}$, and letting the star be solar-like, yields a mass ratio of $\sim$ 100. Therefore, we would expect a qualitatively different fallback rate -- owing to the inability of self-gravity to act quickly enough -- for mass ratios $\lesssim$ 100.}} The top panel of Figure \ref{fig:fb} plots the evolution of the predicted fallback rates over time calculated with the 128 million particle $dM/d\epsilon$ curves. As the time at which the $dM/d\epsilon$ curve is measured shifts to later times, the peak fallback rate increases and occurs earlier while the asymptotic limit of the fallback rate decreases; these two trends are not independent, as a higher peak at earlier times means that less mass will return to the black hole at later times. At very late times this trend reverses slightly, and the fallback rate remains effectively unchanged if the $dM/d\epsilon$ curve at times greater than $\gtrsim 630$ minutes is used. To further illustrate this behavior, the bottom panel of Figure \ref{fig:fb} shows the peak fallback rate as a function of the time post-disruption at which the peak is predicted to occur. The color indicates the time after pericenter when the $dM/d\epsilon$ curve was used to calculate the fallback rate, with early times in blue and later times in red. The fallback rate predicted under the frozen-in approximation closely aligns with the fallback rate calculated at pericenter, meaning it underestimates the peak fallback rate and overestimates the asymptotic limit shortly after pericenter.

\subsection{Fourier analysis}
\label{sec:fourier}
It is apparent by eye from Figure \ref{fig:100M_dmde} that there is, at early times (between $\sim 20 - 30$ minutes post-disruption), a sinusoidal and physical oscillation along the $dM/d\epsilon$ curve with a wavelength of $\sim \Delta \epsilon/8$. To more rigorously assess the strength of this oscillatory feature, we Fourier transform the $dM/d\epsilon$ curve from the 128 million particle simulation from $-\Delta \epsilon/2$ to $\Delta \epsilon/2$, such that the Fourier coefficients are
\begin{equation}
   c_{\rm n} = \frac{1}{\Delta \epsilon}\int_{-\Delta \epsilon/2}^{\Delta \epsilon/2}\frac{dM}{d\epsilon}e^{-\frac{2\pi i n \epsilon}{\Delta \epsilon}}d\epsilon.
\end{equation}

\begin{figure}
    \centering
    \includegraphics[width=0.47\textwidth]{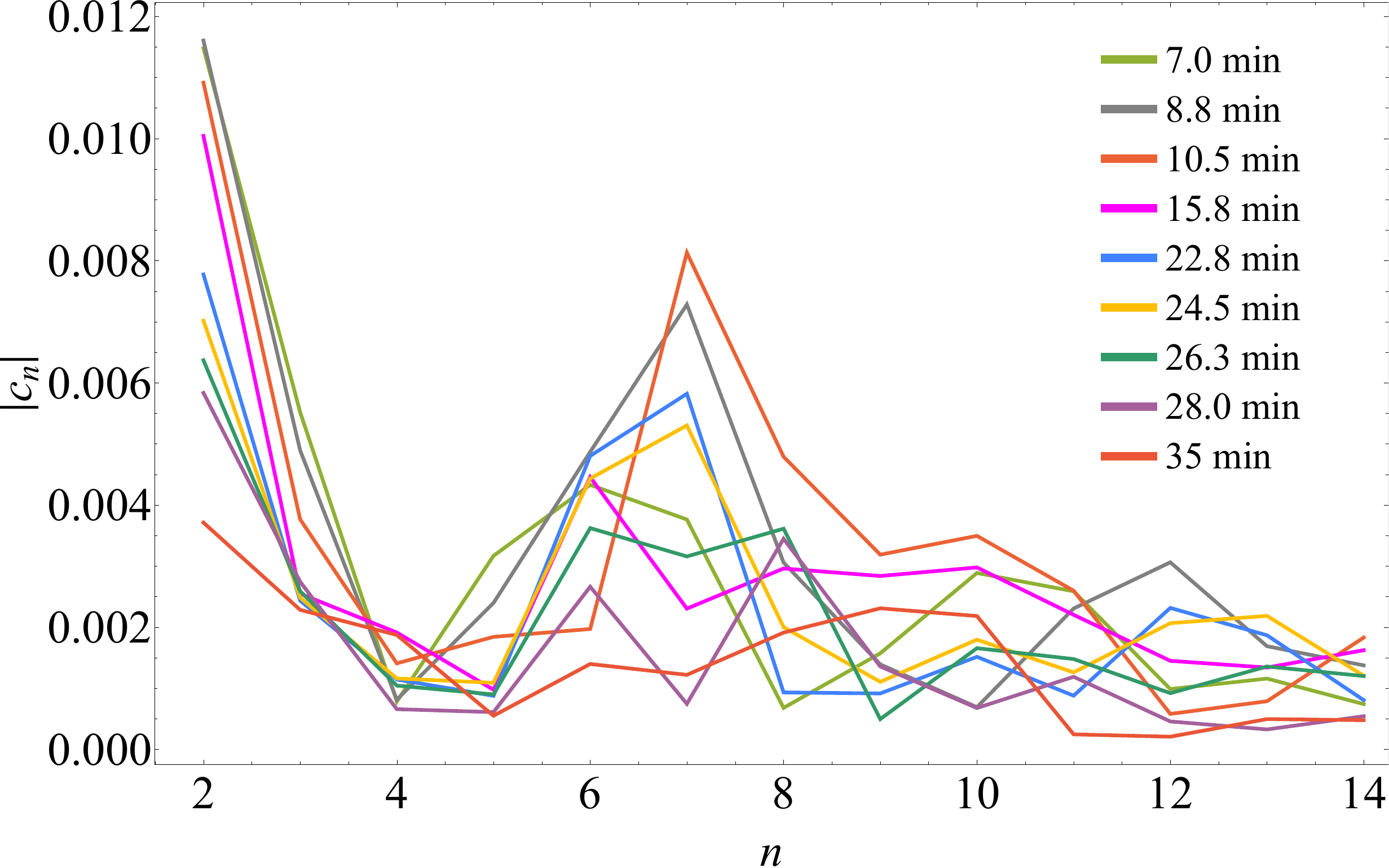}
    \caption{The Fourier coefficients calculated from the $dM/d\epsilon$ curves at the times shown in the legend. The peaks at $n = 6-8$ displayed by some of the curves indicate enhanced power at wavelengths $\simeq \Delta\epsilon/n$.}
    \label{fig:fourier_coeffs}
\end{figure}

Figure \ref{fig:fourier_coeffs} illustrates the Fourier coefficients -- normalized by $c_0$ to offset the declining amount of mass between $-\Delta\epsilon/2$ and $\Delta\epsilon/2$ -- from $n = 2-14$ for the times shown in the legend (the colors match those in Figures \ref{fig:1M_dmde} and \ref{fig:100M_dmde} for the same times); note that there are points only at integer values of $n$, and the lines connecting the points are drawn simply to guide the reader. The coefficients with the largest amplitudes are $n = 0$, which yields the (time-dependent) total mass from $-\Delta \epsilon/2$ to $\Delta \epsilon/2$, and $n = 1$, which accounts for the quadratic-like peak displayed by the $dM/d\epsilon$ curves (especially pronounced at the earliest times). We therefore omitted these values of $n$ from the plot to enhance its readability. This figure demonstrates that, for many of the times indicated in the legend, there is significant and enhanced power at Fourier numbers $n = 6-8$, or wavelengths of $\lambda \simeq \Delta \epsilon/(6-8)$. However, the power is not static: the largest coefficients occur around times of $\sim 10$ and $\sim 25$ minutes, with a relative minimum near $\sim 16$ minutes, and the feature is almost entirely absent by $\sim 35$ minutes.  

\begin{figure*}
    \includegraphics[width=0.485\textwidth]{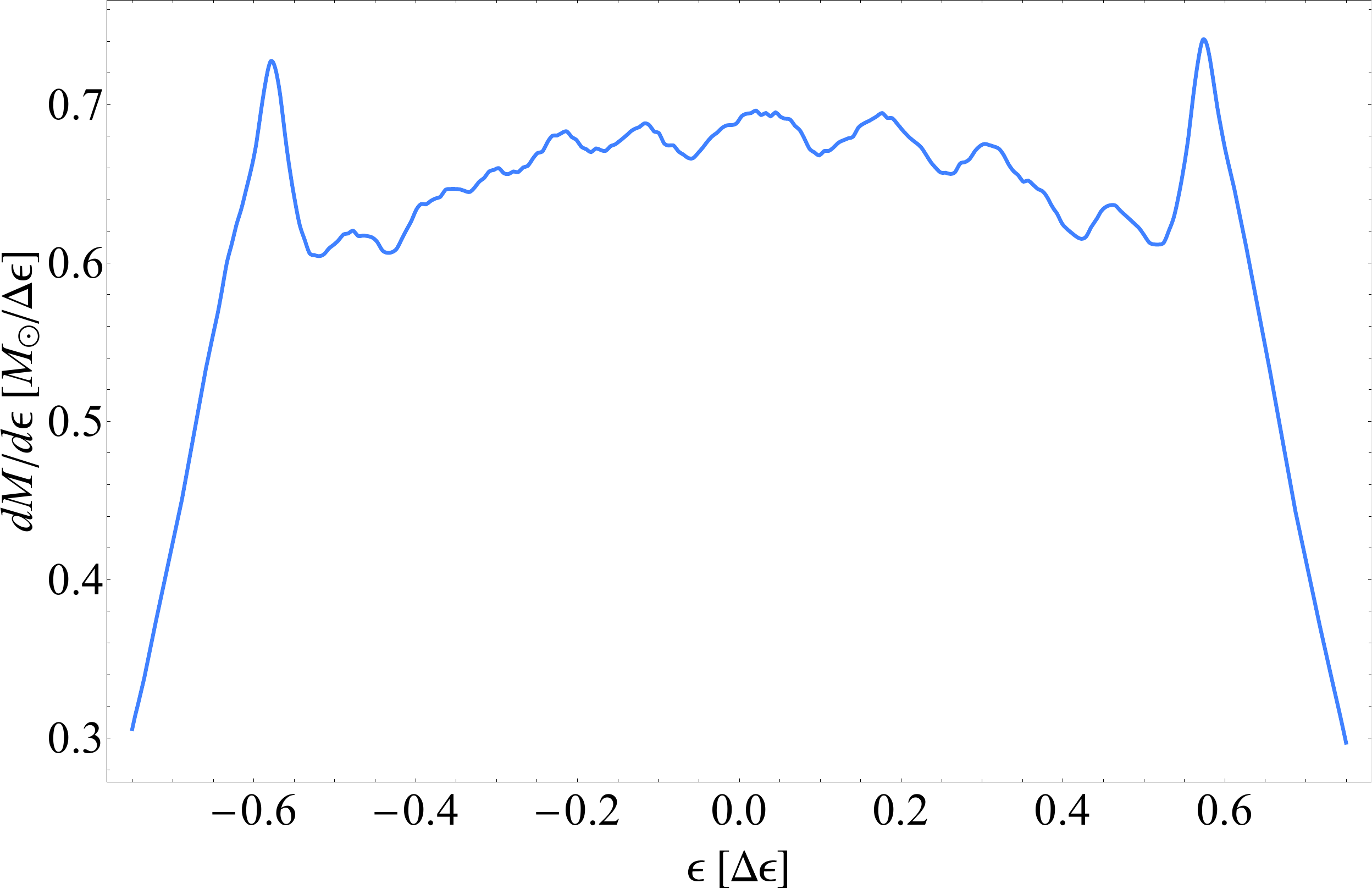}
     \includegraphics[width=0.505\textwidth]{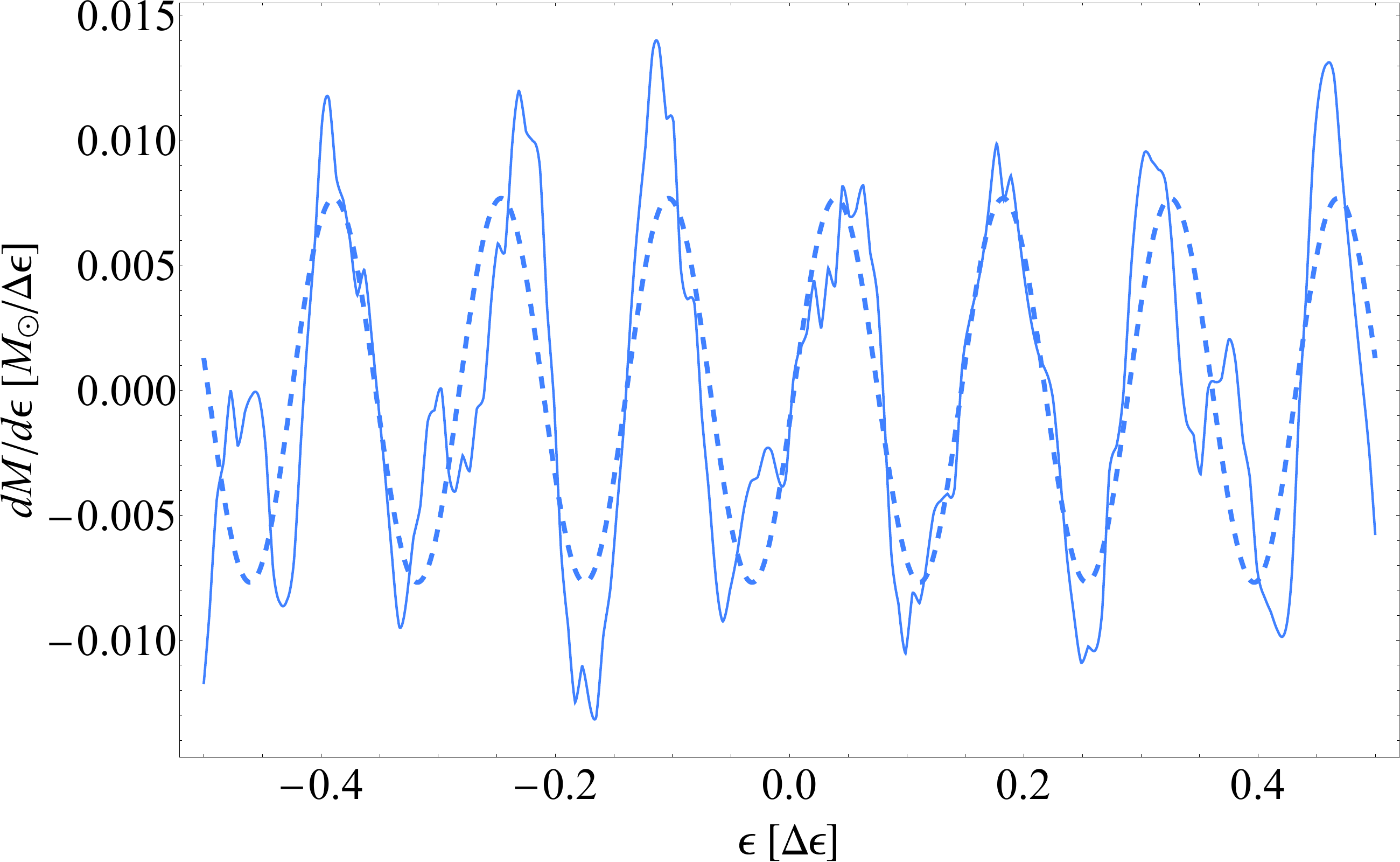}
    \includegraphics[width=0.485\textwidth]{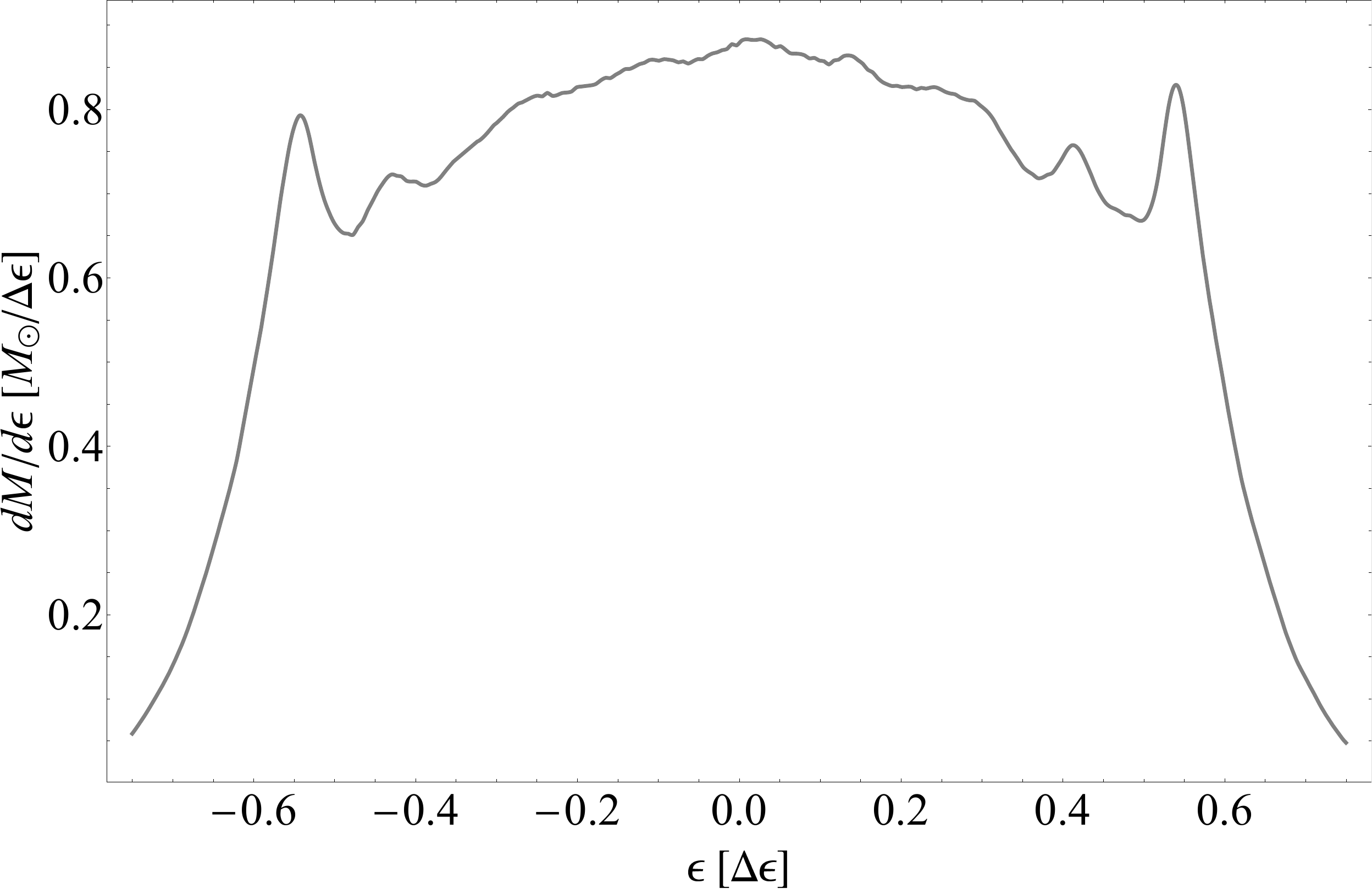}
     \includegraphics[width=0.505\textwidth]{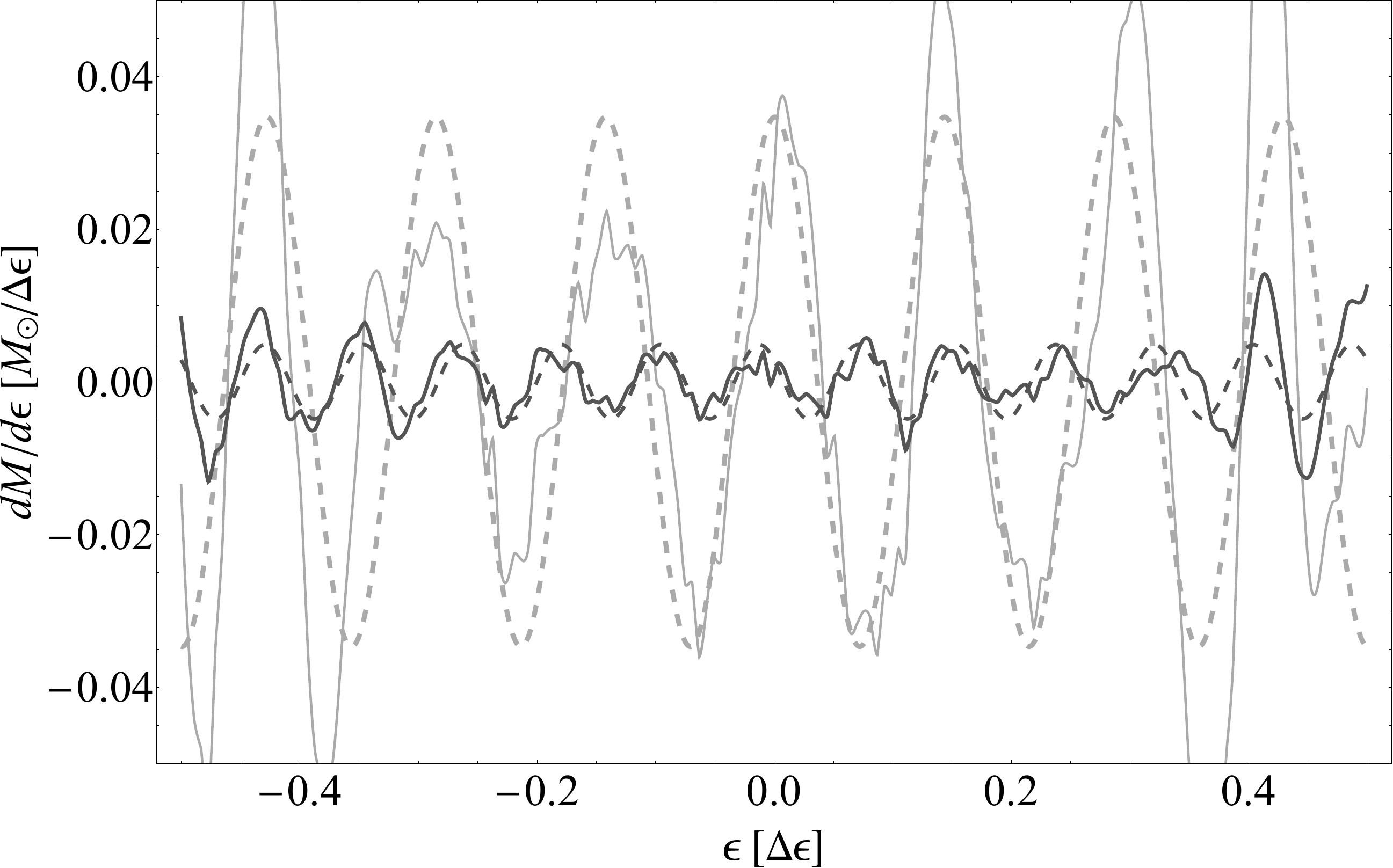}
     \caption{Top: The $dM/d\epsilon$ curve at a time of 22.8 minutes post-pericenter (left), and the high-passed $dM/d\epsilon$ curve -- where the first 6 Fourier modes are removed (right). The  dashed curve in the right panel shows the $n = 7$ term in the Fourier decomposition, i.e., a sinusoidal curve with a wavelength $\Delta\epsilon/7$. Bottom: The $dM/d\epsilon$ curve at a time of 8.8 minutes post-pericenter (left), and the high-passed $dM/d\epsilon$ curve (right); the solid and light-gray curve shows the high-passed solution with the first 6 Fourier modes removed, while the solid and dark-gray curve shows the high-passed data with the first 11 modes removed. The dashed curves show the 7th (light-gray) and 12th (dark-gray) terms in the Fourier expansion. To isolate the curves and increase the clarity of the data, we multiplied the light-gray curves by a factor of 3.}
     \label{fig:highpass}
\end{figure*}

To further highlight the presence of the sinusoidal variation imprinted on the energy distribution, one can high-pass filter the data by subtracting from the $dM/d\epsilon$ curve the first $n$ Fourier modes modes preceding the peak in the power spectrum. The top panel of Figure \ref{fig:highpass} illustrates this technique for the $dM/d\epsilon$ curve at $t = 22.8$ minutes post-disruption. In particular, the top-left panel shows the $dM/d\epsilon$ curve at that time, while the right panel shows the same curve but with the first 6 Fourier modes removed; the dashed curve in the top-right panel illustrates the 7th term in the Fourier series, highlighting the agreement between this Fourier mode -- which has a wavelength of $\Delta \epsilon/7$ -- and the sinusoidal variation of the high-passed data.

\begin{figure}
    \centering
    \includegraphics[width=0.475\textwidth]{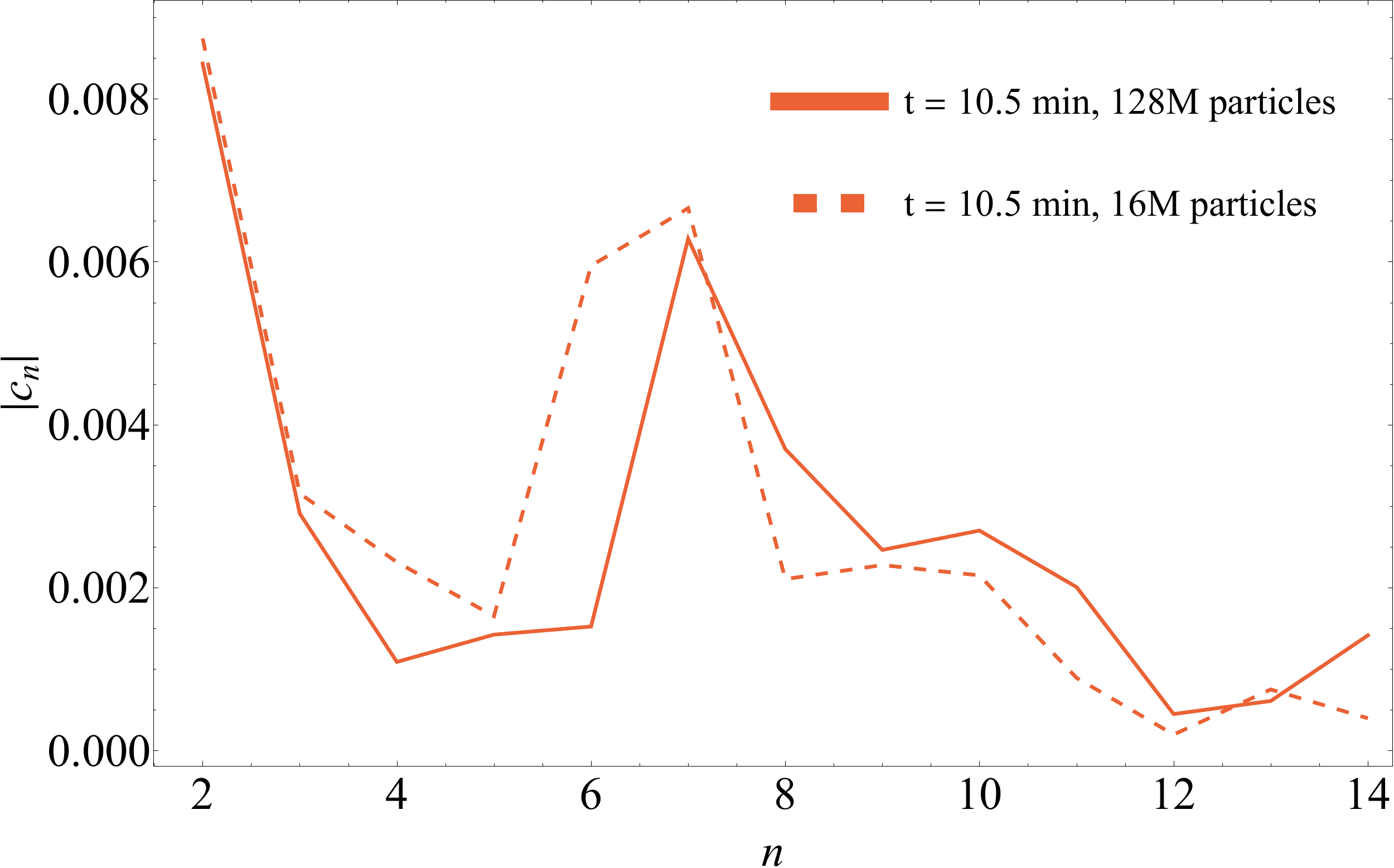}
    \caption{The power spectrum at a time of $10.5$ minutes post-disruption, calculated from the $dM/d\epsilon$ curve with 128 million particles (solid) and 16 million particles (dashed). This highlights the resolution-independence of the excess power at a wavelength in energy of $\sim \Delta \epsilon/7$.}
    \label{fig:fourier_t10p5}
\end{figure}

In addition to the prominent peak in the power spectrum around a wavelength of $\Delta\epsilon/7$, Figure \ref{fig:fourier_coeffs} also shows that there is nontrivial power at higher frequencies, e.g., the solution at a time of 8.8 minutes post-disruption that has a second prominent peak at $n = 12$. The bottom-left panel gives the $dM/d\epsilon$ curve at this time, while the light-gray curves in the bottom-right panel show the high-passed data up to $n = 6$ (solid) and the $n = 7$ Fourier mode (dashed). We scaled the light-gray curves by a factor of 3 to help distinguish them from the dark-gray curves, which show the high-passed data up to $n = 11$ (solid) and the Fourier mode with $n = 12$ (dashed). This shows that, while neither mode is immediately apparent in the $dM/d\epsilon$ curve itself, the sinusoidal variations at these wavelengths are highly significant\footnote{Note that the number of bins used to generate the $dM/d\epsilon$ curves is 500, meaning that the Nyquist frequency corresponds to $n \simeq 500/6 \simeq 83$, and hence we are analyzing excess power at frequencies well below the values at which we are limited by our bin size.}. This significance is further substantiated by Figure \ref{fig:fourier_t10p5}, which shows the power spectrum at $t = 10.5$ minutes from the 128M-particle run (solid) and the 16M-particle run; enhanced power at $n \simeq 7$ is obvious in both cases.

\section{Summary and Conclusions}
When a star is destroyed by the gravitational field of a black hole in a TDE, a useful analytical approach for understanding the ensuing dynamics -- known as the frozen-in approximation and originally due to \citet{lacy82} -- assumes that the gas evolves purely ballistically in the gravitational field of the black hole. Under this approximation, the specific Keplerian energy is a conserved Lagrangian (i.e., per-fluid-element) quantity, implying that the binding energy distribution, $dM/d\epsilon$, is unaltered from the time of disruption. It was shown by \citet{lodato09} that this energy distribution, and the corresponding return rate of gas to the black hole, can be calculated analytically from the density profile of the star. \citet{lodato09} also ran smoothed-particle hydrodynamical simulations to test their analytical model and found noticeable discrepancies, among those being the presence of ``wings'' in the numerically obtained $dM/d\epsilon$ curve that were absent from the frozen-in prediction. \citet{lodato09} attributed these features to one physical effect missing from the frozen-in approximation but included in their hydrodynamical simulations, namely pressure in the form of shocks that putatively occurred near pericenter.

However, there is a second mechanism that can modify the energy distribution of the debris and that was included in the simulations of \citet{lodato09}, namely the self-gravity of the gas. To investigate which of these two effects -- pressure or self-gravity -- is predominantly responsible for the presence of the features in the $dM/d\epsilon$ curve that are not predicted by the frozen-in method, we ran two sets of hydrodynamical simulation of the disruption of a $\gamma = 5/3$, polytropic and solar-like star by a $10^6 M_{\odot}$ black hole (identical in setup to one of the simulations ran by \citealt{lodato09}). In one set of simulations, self-gravity was turned off after the stellar center of mass reached pericenter (equal to the canonical tidal radius). In this case, the specific energy distribution of the debris was effectively unchanged (i.e., showed little to no temporal evolution) from the time of pericenter onward and was in overall good agreement with the frozen-in prediction, as shown in the right panel of Figure \ref{fig:1M_dmde}. Therefore, pressure alone is dynamically insignificant -- the specific Keplerian energy is a conserved Lagrangian variable -- and shocks near pericenter are not responsible for producing the deviation in the $dM/d\epsilon$ curve from the frozen-in prediction. 

In a second set of simulations that varied in resolution, from 250 thousand (comparable to the highest resolution employed in \citealt{lodato09}) to 128 million particles, the self-gravity of the gas was included for the duration of each simulation. In these simulations and as shown in the left panel of Figure \ref{fig:1M_dmde}, the $dM/d\epsilon$ curve showed significant temporal evolution, flattening substantially from the initial distribution (which matched that of the pressure-only simulation at pericenter) in the first $\sim 30$ minutes (roughly the dynamical time of the original star) after pericenter was reached. At later times a central peak reformed, with the evolution of the debris stream slowing substantially by $\sim few$ days following pericenter, after which time it agreed with the numerical results of \citet{lodato09}. We therefore conclude that self-gravity -- and in particular the component of the gravitational field that acts along the direction of the stream (see also \citealt{steinberg19}, who similarly argued that the influence of self-gravity along the stream is responsible for the time-dependent modification of the energy spread during the ingress of the star through the tidal sphere) -- is the physical mechanism responsible for the evolution of the energy distribution of the debris\footnote{There is also an asymmetry in the peaks that form around $\pm 0.5\Delta\epsilon$, with the peak at $+0.5\Delta\epsilon$ systematically higher than that at $-0.5\Delta\epsilon$, which is particularly noticeable in the high-resolution simulations in Figure \ref{fig:overlay}; we thank Matt Todd for pointing this out to us. This asymmetry is likely due to the fact that the shear in the unbound segment of the stream is reduced compared to that in the bound half \citep{coughlin16}, rendering self-gravity more capable of modifying the mass distribution.}, and shocks are irrelevant.

Self-gravity results in clear sinusoidal variations in the specific energy distribution shortly after the star reaches pericenter; these are apparent by-eye in Figure \ref{fig:100M_dmde} around $\sim 20$ minutes post-pericenter, and a Fourier analysis of the stream shows increased power on energy scales of $\sim \Delta \epsilon/7$ (see Figures \ref{fig:fourier_coeffs} and \ref{fig:highpass}). On the one hand, increased power at a specific \emph{spatial} scale is not unexpected, as hydrostatic cylinders are gravitationally unstable with a fastest-growing mode that peaks at a wavenumber of $\sim 1.2 H$ for a $5/3$-polytropic gas \citep{coughlin20}, where $H$ is the cross-sectional radius of the cylinder. If the specific Keplerian energy were exactly conserved, then the energy would correlate with the initial position of a Lagrangian fluid element, and thus we would expect pronounced power on the energy scale that correlates with the length scale of the most unstable mode. Since the Keplerian energy is not actually a conserved Lagrangian variable (as we have shown in the present analysis) this conclusion is only tentative, but it is suggestive of the fact that this oscillation in energy space is a result of the gravitational instability of the stream. The details (e.g., the growth rate, the dispersion relation) of the instability are significantly more complicated than the hydrostatic case owing to the presence of the shear along the filament axis, but the recent work of \citet{coughlin23} has shown that the stream is overstable in the cylindrical-radial direction, and it is plausible that the oscillation in energy space identified here is a manifestation of an analogous instability along the filament axis.

\section*{Acknowledgements}
We thank Matt Todd for a useful comment regarding the asymmetry of the mass distribution in the stream, and we thank the referee, Elad Steinberg, for useful comments that added to the quality of our work. JF acknowledges support from Syracuse University through the SOURCE program and a Young Research Fellows award. ERC acknowledges support from the National Science Foundation through grant AST-2006684, and the Oakridge Associated Universities for their support through a Ralph E.~Powe Junior Faculty Enhancement Award. CJN acknowledges support from the Science and Technology Facilities Council (grant number ST/Y000544/1), and the Leverhulme Trust (grant number RPG-2021-380). Part of this research made use of the DiRAC Data Intensive service at Leicester, operated by the University of Leicester IT Services, which forms part of the Science and Technology Facilities Council (STFC) DiRAC HPC Facility (www.dirac.ac.uk). The equipment was funded by BEIS capital funding via STFC capital grants ST/K000373/1 and ST/R002363/1 and STFC DiRAC Operations grant ST/R001014/1. DiRAC is part of the National e-Infrastructure.

%%%%%%%%%%%%%%%%%%%%%%%%%%%%%%%%%%%%%%%%%%%%%%%%%%
\section*{Data Availability}

The simulations in this paper can be reproduced by using the \textsc{phantom} code (Astrophysics Source Code Library identifier ascl.net/1709.002). The data underlying this article will be shared on reasonable request to the corresponding author.



\bibliographystyle{mnras}
%\bibliography{refs} 

% Don't change these lines
\bsp	% typesetting comment
\label{lastpage}
\end{document}